\documentstyle{article}
\begin{document}

\centerline{\bf {Abstract.}}

\newcommand{\h}{\LARGE}
\newcommand{\smc}{} 
\newcommand{\bier}{} 

 \vspace{-15mm}

{\bf {
\centerline{ A NEW APPROACH TO QUANTIZATION OF GRAVITY. }
\centerline{  2+1-DIMENSIONAL EXAMPLE. }}}
\centerline{ }
\centerline{ \it S.N.Vergeles.}
\vspace{3mm}
\centerline { {\it Russian Academy of Sciences. } }
\centerline { {\it The Landau Institute for Theoretical Physics, Moscow,
 Russia.}}

  \parbox[b]{135mm}{
\baselineskip=14pt
 { \hspace{5mm}
In this paper the quantization of the 2$+$1-dimensional gravity
couplet to the massless Dirac field is carried out. The problem is solved by
the application of the new Dynamic Quantization Method [1,2].
 It is well-known that in general
covariant theories such as gravitation, a Hamiltonian is any linear
combination of the first class constraints, which can be considered
as gauge transformation generators.
To perform quantization, the Dirac field modes with gauge invariant
 creation and annihilation operators are selected.
 The regularization of the theory is made by imposing an infinite set
 of the second class constraints: almost all the gauge invariant creation and
 annihilation operators (except for a finite number) are put equal to zero.
 As a result the regularized theory is gauge invariant.
  The gauge invariant
 states are built by using the remained gauge invariant fermion creation
 operators
 similar to the usual construction of the states in any Fock space.
The developed dynamic quantization method can construct a
 mathematically correct perturbation theory in a gravitational constant.}}

\large {
\baselineskip=18pt

\centerline{ }
\centerline{   \bf  {1. Introduction.}}
\centerline{ }

\baselineskip=18pt
   In  the recent paper of  the  author  [2] the Hamilton
Quantization of the 2$+$1-dimensional gravity
couplet to massless Dirac field is carried out on the basis of the Dynamic
Quantization Method. The space of the regularized states of the system is
built. The perturbation theory (PT) in a gravitational constant is developed.
Using it the regularized Heisenberg equations are solved.
Lately some axiomatization of the dynamic quantization method has been
performed using gravity model [2] as an example. This
clarified the inner logics of the dynamic method. It also became possible
to get concrete results easier. This paper considers
the process of dynamic quantization of gravity in three-dimensional
space-time from a new viewpoint. Since the dynamic quantization
method is not well-known, we outline it shortly here.

  Let's consider some field theory. Suppose that in this theory the
physical degrees of freedom in ultraviolet region can be  classified by
the modes with the following properties:

   a) The occupation numbers of the modes are conserved or they are
adiabatic invariants of motion.

   b) The corresponding creation  \ {\mbox {\h $a^+_N$}} \ and annihilation
    \ {\mbox {\h $a_N$}} \ operators are to be gauge invariants.

  These are basic points for the Dynamic Quantization Method.

  Then the regularization of the theory is made by imposing the second class
constraints
\vspace{-15mm}\\
{\h {
$$       a^+_N \approx 0, \qquad a_N\approx 0     \eqno(1.1)         $$
}}
 for quantum numbers  \ {\mbox {\h $N$}} \ from the ultraviolet tail.
 The second class constraints
lead to the substitution of the corresponding Dirac commutational relations
 (CR) for the initial or formal CR. Then we are to solve the Heisenberg
equations obtained by the Dirac CR. Thus
{\it {the theory becomes regularized
with definite numbers of physical degrees of freedom. Moreover, since the
 annihilation and creation operators in Eqs.(1.1) are gauge invariant the
 imposition of the constraints (1.1) does not violate gauge invariance.}}

   It is necessary to pay attention to the difference
   between the Feynman and Dynamic quantizations.
 The Feynman quantization is based on the hypothesis that the interaction
 can be switched on and off  adiabatically. This hypothesis is equivalent
 to the assumption that the physical vacuum differs slightly from the naive
one \\
 ( at a switched off interaction). The assumption leads to the Feynman rules.
 But the Feynman PT in general cannot be regularized so
 that the number of physical degrees of freedom would be definite.
 As Gribov noted [3] this nonconservation of the whole number of degrees
 of freedom results in the appearance of the gauge anomaly.

 It is known, that the Feynman rules admit the Wick rotation which results in
 equivalence between  $(D-1)+1$-dimensional quantum field theories and
 $D$-dimensional classic statistic models. Thus, the Euclidean quantization is
 automatically equivalent to the Feynman one.

  The available experience of the general relativistic theory shows that the
  gravitating universe evolves from peculiarity to peculiarity. Thus, it
  is no wonder that the Feynman theory is not applicable for quantization of
  gravity. On the contrary, the dynamic quantization
  corresponds naturally to the situation in a general relativistic theory.
  Indeed,
  the annihilation and creation operators (see (1.1))
  are gauge invariant and the corresponding field modes carry all space-time
  information about universe evolution; the space-time peculiarities
  are not excluded. These modes satisfy equations of motion, so that the
  corresponding annihilation and creation operators are conserved and
  the imposition of the constraints (1.1) is in agreement with dynamics.
  Thus dynamic quantization is realized. We think that in this case different
  gauge anomalies are absent. Such approach is used
  to investigate gauge anomaly in papers [1,4].

  In this paper the dynamic quantization theory of gravitation interacting
  with the Dirac massless field in 2$+$1$D$ space is present. To perform
  dynamic quantization, it is necessary to have a complete set of gauge
  invariant one-fermion states,  mutually orthogonal in a natural
  sense.
  In terms of these states the regularization and vacuum feeling is produced.
  In the previous work [2] these states appeared as a
result of long constructions.
  In this paper we assume that the necessary
  set of one-fermion states exists in the theory. This assumption can
  simplify essentially theory exposition and make a lot of important
  statements clear. The assumption is
  justified since on its basis  the  selfconsistent,  physically
  sensible and
  mathematically correct theory is built. As it shown in [2] the
  space of the regularized states of the system is constructed, the regularized
  Heisenberg equations for field operators are written out and solved by the
  mathematically correct PT in a gravitational constant.

\baselineskip=18pt

      Let us emphasize that 2$+$1$D$ gravitation theory is close to some
      extent
 to the  chiral  Schwinger  model, studied  earlier  by
dynamic quantization
 method [1]. Indeed, in both theories the gauge fields do not have their own
 local degrees of freedom ( the absence of photon in 1$+$1 $D$ and graviton
 in 2$+$1 $D$).

\baselineskip=18pt
\centerline { }
\centerline {{\bf 2. The Formal Quantization.}}
\centerline { }

\baselineskip=18pt
    Let   \ {\mbox {\h $x^{\mu}=(x^0,x^1,x^2)$}} \  denote local
    coordinates in some 3-Dimensional metric space.
    The metrics is expressed in the form
{\h {
 $$ ds^2=g_{\mu\nu}\,dx^{\mu}dx^{\nu}=
     e^a_{\mu}\,e_{a\nu}\,dx^{\mu}\,dx^{\nu}     \eqno(2.1)   $$
     }}
where \ {\mbox {\h $e^a_{\mu}$}} \ are the dreibein.
   Latin letters \ {\mbox {\h $a,b,\ldots = 0,1,2$}} \  from the
   beginning of \ the alphabet \ are
 \ {\mbox {\h $SO(2,1)$}} \ indices, the sign convention is \\
  \ {\mbox {\h $\eta_{ab}=$}}
   \ {\mbox {\h $diag \,(1,-1,-1)$}} .
  Let \ {\mbox {\h $g^{\mu\nu}$}} \  be the inverse metric tensor and
  {\h {
$$ e^{\mu}_a=g^{\mu\nu}\,\eta_{ab}\,e^b_{\nu}
$$
}}
Then
\vspace{-15mm}\\
{\h {
$$ e^a_{\mu}\,e^{\mu}_b=\delta^a_b, \qquad e^{\mu}_a\,e^a_{\nu}=
 \delta^{\mu}_{\nu}     \eqno(2.2)
$$
}}
The connection 1-form is designated as follows:
 \ {\mbox {\h $ \omega^a_b=\omega^a_{b\mu}\,dx^{\mu}       $}}. Since
 {\h {
 $$ \omega_{ab\mu}=\eta_{ac}\,\omega^c_{b\mu}=-\omega_{ba\mu}  \eqno(2.3) $$
 }}
 it is convenient to use the equivalent quantity [5]
 {\h {
$$ \omega^c_{\mu}=
 \frac{1}{2}\,\varepsilon^{abc}\,\omega_{ab\mu}, $$
$$
\varepsilon^{012}=1      \eqno(2.4)   $$
}}
 In these notations the scalar curvature is of the form
 {\h {
$$ \sqrt g\,R
 =\varepsilon^{\mu\nu\lambda}\,e_{c\lambda}(\partial_{\mu}\omega^c_{\nu}-
    \partial_{\nu}\omega^c_{\mu}-
   \varepsilon_{ab}{ }^c\,\omega^a_{\mu}\,\omega^b_{\nu})  $$
$$
     g=det\,g_{\mu\nu}                  \eqno(2.5)    $$
}}

    Let  \ {\mbox {\h $\gamma^a$}} \  be the Dirac matrices
     \ {\mbox {\h $2\times 2$}} \ which satisfy
    the conditions
    {\h {
 $$  \gamma^a\gamma^b+\gamma^b\gamma^a=2\,\eta^{ab}   \eqno(2.6)      $$
 }}
  Denote by \ {\mbox {\h $\psi$}} \
 and \ {\mbox {\h $\bar {\psi}=\psi^+\gamma^0 $}} \
 Dirac two-component complex field. Cross above means Hermithian
conjugation.

    According to (2.5) the simplest general covariant action is of the form
   {\h {
$$
 A=\int\,d^3x\,\{-\frac{1}{16\pi G}\,\varepsilon^{\mu\nu\lambda}\,e_{c\lambda}
(\partial_{\mu}\,\omega^c_{\nu}-\partial_{\nu}\,\omega^c_{\mu}-
 \varepsilon_{ab}{ }^c\omega^a_{\mu}\omega^b_{\nu})+   $$
$$   + \frac{i}{2}(\bar {\psi}\,\Gamma^{\mu}\,\nabla_{\mu}\psi-
 \nabla_{\mu}\bar {\psi}\,\Gamma^{\mu}\,\psi)\},           \eqno(2.7)
$$
}}
 where
 {\h {
$$
  \nabla_{\mu}\psi= (\partial_{\mu}-
 \frac{i}{2}\,\omega_{a\mu}\gamma^a)\psi, \qquad
 \nabla_{\mu}\bar {\psi}=\partial_{\mu}\bar {\psi}+
 \frac{i}{2}\,\bar {\psi}\,\gamma^a\omega_{a\mu},   \eqno(2.8)
$$
}}
and
{\h {
$$
\Gamma^{\mu}=\sqrt g\,e^{\mu}_a\,\gamma^a=
 \frac{1}{2}\,\varepsilon^{\mu\nu\lambda}\,
  \varepsilon_{abc}\,e^b_{\nu}\,e^c_{\lambda}\,\gamma^a       \eqno(2.9)
$$
}}

   Note  that  the  Fermi  part  of  the   action   (2.7)   is   taken
symmetrised
in order the action to be Hermithian. Indeed the expression
{\h {
 $$
 \int\,d^3x\,i\,\bar\psi\,\Gamma^{\mu}\nabla_{\mu}\psi
 $$
 }}
 is not Hermithian if the connection has torsion. But the equations of
 motion following from the action (2.7) lead to the connection
 with the torsion.

    Further we shall designate by the letters \ {\mbox {\h $x,\,y,\ldots ,$}} \
   the totality of spatial
   coordinates  \ {\mbox {\h $(x^1,x^2),\,(y^1,y^2),\ldots,$}} \
 the time
 argument \ {\mbox {\h $x^0$}} \  will be omitted and the point above
 will mean the derivative \ {\mbox {\h $\partial/\partial x^0 $}} \ and
 {\h {
 $$
 d^2x=dx^1\,dx^2, \ \qquad \ \delta^{(2)}(x)=\delta(x^1)\,\delta(x^2)  .
 $$
 }}

\baselineskip=18pt
   When obtaining Hamiltonian from the action (2.7) the following difficulty
 arises. Since the action (2.7) contains the quantities
 \ {\mbox {\h $\dot {\psi}$}} \ and
 \ {\mbox {\h $\dot {\bar {\psi}}$}} \ , the fields
 \ {\mbox {\h $\psi$}} \ and \ {\mbox {\h $\bar {\psi}$}} \  both are to be
 regarded as coordinate variables. But the corresponding momentum variables
 \ {\mbox {\h $\pi_{\psi}$}} \ and \ {\mbox {\h $\pi_{\bar {\psi}}\,$}} \ are
 expressed through the same fields
 \ {\mbox {\h $\psi$}} \ and \ {\mbox {\h $\bar {\psi}$}} \ .
 Therefore the constraints
\vspace{-11mm}\\
 {\h {
 $$
 \bar {\tau}=\pi_{\psi}-\frac{1}{2}\,\bar {\psi}\,\Gamma^0 \approx 0 \ , \qquad
 \tau=\pi_{\bar {\psi}}-\frac{1}{2}\,\Gamma^0\psi \approx 0   \eqno(2.10)
 $$
 }}
 take place. Thus the Hamiltonian dynamic variables unite in the following
 pairs:
 {\h {
 $$
 (\psi,\pi_{\psi}), \qquad \ (\bar {\psi},
 \pi_{\bar {\psi}}),  \qquad \  (\omega^a_i,{\cal P}^i_a), $$
 $${\cal P}^i_a=-\frac{1}{8\pi G}\,\varepsilon_{ij}\,e_{aj}    \eqno(2.11)
 $$
 }}
\baselineskip=18pt
 Latin letters \ {\mbox {\h $i,\,j,\,\ldots =1,2\,$}} \  from the middle
 of the alphabet are spatial indices, \ {\mbox {\h $\varepsilon_{ij}
 =-\varepsilon_{ji}, \ \varepsilon_{12}=1$}} . The
 fields   \ {\mbox {\h $\omega^a_0=\omega^a$}} \  and
 \ {\mbox {\h $e^a_0=e^a$}} \  play
 the role of the Lagrange multiplier.

\baselineskip=18pt
   Let us introduce the function \ {\mbox {\h $\alpha$}} \  defined on
   the homogeneous operators
 with the values in the group {\smc Z}  {\mbox {\h ${ }_2$}}. By definition
 {\h {
 $$
 \alpha (\omega^a)=\alpha (e^a)=\alpha (\omega^a_i)=
 \alpha ({\cal P}^i_a)=0, $$
$$
 \alpha (\psi)=\alpha (\pi_{\psi})=\alpha (\bar {\psi})=
 \alpha(\pi_{\bar {\psi}})=1
 $$
 }}
    If the function \ {\mbox {\h $\alpha$}} \ is defined on the operators
 \ {\mbox {\h $A$}} \ and \ {\mbox {\h $B$}} \ ,
 then \ {\mbox {\h $\alpha (AB)=$}} \\
 {\mbox {\h $\alpha (A)+\alpha (B) (mod\,2)$}} \ .
 Everywhere we understand under the commutator of the gomogeneous operators
 \ {\mbox {\h $A$}} \ and \ {\mbox {\h $B$}} \  the expression
 {\h {
 $$
  [A,B]=AB-BA\,(-1)^{\alpha (A)\alpha (B)}           \eqno(2.12)
 $$
 }}
 The initial nonzero simultaneous CR are of the form
 {\h {
 $$
 [\omega^a_i(x),{\cal P}^j_b(y)]=
 i\,\delta_{ij}\,\delta^a_b\,\delta^{(2)}(x-y), $$
$$
 [\psi(x),\pi_{\psi}(y)]=1\cdot \delta^{(2)}(x-y),  $$
$$
 [\bar {\psi}(x),\pi_{\bar{\psi}}(y)]=1\cdot \delta^{(2)}(x-y) \eqno(2.13)
 $$
}}
   The Hamiltonian of the system looks like:
{\h {
 $$
 H=H_{\chi}+H_{\phi},  $$
$$
 H_{\chi}=-\int\,d^2x\,\omega^a\chi_a,    $$
$$
 H_{\phi}=\frac{1}{8\pi G} \int\,d^2x\,e_a\phi^a, \eqno(2.14a)  $$
 }}
 where
 {\h {
$$
 \chi_a=\partial_i{\cal P}^i_a-
  \frac{1}{2}\,\varepsilon_{ab}{ }^c\,(\,\omega^b_i\,{\cal P}_c^i+
 {\cal P}_c^i\,\omega^b_i\,)+
$$
$$
$$
$$
 + \frac{1}{2}\,\bar {\psi}\,\sqrt g\,e^0_a\,\psi,   \eqno(2.14b)  $$
$$
$$
$$
 \phi^a=\frac{1}{2}\,\varepsilon_{ij}(\partial_i\omega^a_j-
 \partial_j\omega^a_i-\varepsilon^a{ }_{bc}\,\omega^b_i\,\omega^c_j)+   $$
$$
$$
 $$ +\frac{i}{2}\,(8\pi G)^2\varepsilon_b{ }^{ca}
 (\bar {\psi}\,\gamma^b\,{\cal P}^i_c\,\partial_i\psi-
 \partial_i\,\bar {\psi}\,{\cal P}^i_c\,\gamma^b\,\psi)+$$
$$
$$
$$
 +\frac{1}{4}\,(8\pi G)^2\varepsilon_b{ }^{ca}\,\bar {\psi}
 \{{\cal P}^i_c\,\omega_{di}\,[\,s\,\gamma^b\gamma^d+(1-s)\,\gamma^d\gamma^b]+
 $$
 $$
 $$
 $$ +\omega_{di}\,{\cal P}^i_c\,[\,(1-s)\,\gamma^b\gamma^d+
 s\,\gamma^d\gamma^b\,]\}\,\psi         \eqno(2.14c)
$$
}}
   The quantities \ {\mbox {\h $\sqrt g\,e^{\mu}_a$}} \ are to be expressed
   through Hamiltonian
 variables:
 {\h {
 $$
 \sqrt g\,e^0_a=\frac{1}{2}\,(8\pi G)^2
 \,\varepsilon_a{ }^{bc}\,\varepsilon_{ij}
 \,{\cal P}^i_b\,{\cal P}^j_c,
 $$
 $$
 \sqrt g\,e^i_a=-e_c\,(8\pi G)\,\varepsilon_a
 { }^{bc}\,{\cal P}^i_b  \eqno(2.15)
 $$
 }}
\baselineskip=18pt
   In (2.14c) \ {\mbox {\h $s$}} \ is a free real parameter. The freedom
   in the choice of \ {\mbox {\h $s$}} \
 means the freedom in the arrangement of operators in the Hamiltonian
 (2.14). This freedom is eliminated in the next section.

\baselineskip=18pt
 With the help of CR (2.13) we find the commutator of the constraints (2.10):
{\h {
$$
 [{\bar {\tau}}_{\sigma}(x),\,\tau_{\rho}(y)]=
 -\delta^{(2)}(x-y)\,\Gamma^0_{\rho \sigma}(x)       \eqno(2.16)
 $$
 }}
 It is seen from here that constraints (2.10) are the second class
 constraints. Thus it is necessary to pass from initial CR (2.13) to
 the corresponding Dirac CR. This is made according to general procedure [6].
 Since further the initial CR (2.13) are not used we keep the previous
 notations for the so-obtained Dirac CR.

\baselineskip=18pt
  Write out the nonzero simultaneous Dirac CR for fundamental fields
 \ {\mbox {\h $\omega^a_i, \ {\cal P}^i_a, $}} \\
 {\mbox {\h $ \psi, \ \bar {\psi}$}} \ :
\vspace{-15mm}\\
 {\h {
 $$
  [\psi_{\sigma}(x),\,{\bar {\psi}}_{\rho}(y)]=\delta^{(2)}
(x-y)(\Gamma^0)^{-1}_{\sigma \rho}(y), $$
$$
$$
$$ [{\cal P}^i_a(x),\,\omega^b_j(y)]=-i\,\delta_{ij}\,
 \delta^b_a\,\delta^{(2)}(x-y),  $$
 $$
 $$
$$ [\omega^a_i(x),\,\psi(y)]=
\frac{i}{2}\,\delta^{(2)}(x-y)\,(8\pi G)^2\,$$
$$
$$
$$\varepsilon_{ij}\,\varepsilon_b{ }^{ca}\,
 \bigl(\,{\cal P}^j_c\,(\Gamma^0)^{-1}\,
 \gamma^b\,\psi\,\bigr)\,(y),   $$
 $$
 $$
$$ [\omega^a_i(x),\,\bar {\psi}(y)]=\frac{i}{2}\,\delta^{(2)}(x-y)\,(8\pi G)^2
$$
\vspace{-3mm}
$$ \,\varepsilon_{ij}\,\varepsilon_b{ }^{ca}\,(\,\bar {\psi}
 \gamma^b\,(\Gamma^0)^{-1}\,{\cal P}^j_c\,)\,(y),  $$
 $$
 $$
$$ [\omega^a_i(x),\,\omega^b_j(y)]=\frac{i}{2}\,\delta^{(2)}(x-y)\,(8\pi G)^2
$$
\vspace{-3mm}
$$ \varepsilon_{ij}\,\bigl(\bar \psi\,\frac{e^{0a}\,e^{0b}}{e^0_c\,e^{0c}}
 \psi \,\bigr)\,(y)     \eqno(2.17) $$
 $$
 $$
 }}
 The quantities  \ {\mbox {\h $e^0_a$}} \ are given according to (2.15).

\baselineskip=18pt
  By the direct calculations one can be convinced that CR (2.17) satisfies the
  following properties for any complex numbers \ {\mbox {\h $x, y$}} \ :
  {\h {
 $$
 [A,\,B]=-[B,\,A](-1)^{\alpha (A)\alpha (B)},   $$
 \vspace{-1mm}
$$
 [xA+yB,\,C]=x\,[A,\,C]+y\,[B,\,C],
 $$
 \vspace{-1mm}
 $$
 [A,\,[B,\,C]](-1)^{\alpha (A)\alpha (C)}+
 [B,\,[C,\,A]](-1)^{\alpha (A)\alpha (B)}+
 $$
 \vspace{-1mm}
 $$
 +[C,\,[A,\,B]](-1)^{\alpha(B)\alpha (C)}=0     \eqno(2.18)
 $$
 }}
\vspace{-10mm}\\

\baselineskip=18pt
  Take by definition
\vspace{-14mm}\\
 {\h {
 $$
 [A,\,BC]=[A,\,B]\,C+B\,[A,\,C]\,(-1)^{\alpha (A)\alpha (B)}
  \eqno(2.19)
$$
}}
\vspace{-1mm}\\
 Formulae  (2.17)  -  (2.19)  define  inductively  CR  for  any
functional on fundamental fields
  \ {\mbox {\h $\omega^a_i, \ {\cal P}^i_a, \ \psi, \ \bar {\psi}$}}.

  Thus, in Section 2 the formal quantization of the system is performed:
  Hamiltonian is of  the  form  (2.14)  and  any  CR  are  defined
according to  (2.17) - (2.19).

\baselineskip=18pt

\centerline{ }
\centerline{ }
\centerline {\bf {3. The Separation of the gauge
invariant degrees of freedom }}
\centerline{\bf {and regularization. }}
\centerline { }

\baselineskip=18pt

  Our aim consists in the construction of
  states which are annulled by operators (2.14), and in the solution of
  the Heisenberg
  equations
   \ {\mbox {\h $ i\,\dot{A}=[A,\,H]$}} \ . Here
  \ {\mbox {\h $A$}} \ - is any of fundamental fields
  (or their functional), and \ {\mbox {\h $H$}} \
  is given by (2.14).

      However this problem can be solved only together with the problem of
 selection of gauge invariant degrees of freedom and regularization
 of the theory. The problem is solved in this Section.

\centerline{ }
\centerline{ \it {3.1 The main Hypothesis and its Consequences.}}
\centerline{ }

\baselineskip=18pt
We shall hold the point of view accepted in axiomatic quantum field theory.
 Let us begin at the construction of
 the Fock space formalism for fermion degrees of freedom.
 The space of physical
 states of gravitational field is discussed in the beginning of the next
 Section.

\baselineskip=18pt
 Denote by \ {\mbox {\h $\{\vert \, N\rangle\}$}} \ the complete set of
 one-fermion states and by \ {\mbox {\h $\vert \, 0\rangle$}} \
 the fermion vacuum so that
\vspace{-13mm}\\
 {\h {
  $$
  \psi(x)\,\vert \, 0\rangle =0, \qquad  \psi(x)\,\vert \, N\rangle =
  \psi_N(x)\,\vert \, 0\rangle ,
  \eqno(3.1) $$
  }}
  where \ {\mbox {\h $\psi_N(x)$}} \ is the Bose-operator field.

 Any fermion state is built by  one-fermion  states.
For example,
 two-fermion state is of the form:
 {\h {
 $$ \vert \, N_1\rangle \otimes \vert \, N_2\rangle =
 -\vert \, N_2\rangle \otimes \vert \, N_1\rangle       \eqno(3.2)
 $$
 }}

 It is convenient to modify the previous designations by introducing
 vacuum state for any index {\mbox {\h $N$}}. Then, for any
 \ {\mbox {\h $N$}} \ there are two states
 {\h {
 $$ \vert \, N,\lambda_N\rangle, \qquad  \lambda_N=0,1,
 $$
 $$ \vert \, 0\rangle =\bigotimes_N \vert \, N,0\rangle, \qquad
 \vert \, N,1\rangle \equiv \vert \, N\rangle
 $$
 }}
\baselineskip=18pt
Let us introduce the arrangement in space of index
 \ {\mbox {\h $N$}}. Now any fermion state is expressed as a linear
 combination of the following states:
\vspace{-12mm}\\
 {\h {
 $$
 \vert \, N_1,\ldots ,N_s\rangle=\bigotimes_N \vert \, N,\lambda_N\rangle
\eqno(3.3)
 $$
 }}
 Here \ {\mbox {\h $N_1<N_2<\ldots <N_s$}} \ are indices for which
 \ {\mbox {\h $ \lambda_{N_1}=$}}
 \ {\mbox {\h $\lambda_{N_2}=\ldots$}} \\
  {\mbox {\h $=\lambda_{N_s}=1$}} \ ,
  \ {\mbox {\h $s$}} \ is any natural number.
 The tensor products (3.3) are ordered so that the state
 \ {\mbox {\h $ \vert\,N_i,1\rangle $}} \
 is on the left of \, {\mbox {\h $ \vert\,N_j,1\rangle $}}
 at \ {\mbox {\h $ N_i<N_j $}} \ .

 Define the fermion annihilation operators \ {\mbox {\h $ a_N$}} \
by their action on states (3.3)
\vspace{-14mm}\\
{\h {
$$
a_{N_i}\, \vert \, N_1,\ldots ,
N_s\rangle=(-1)^{\lambda(N_1,\ldots ,N_s;\,N_i)}\cdot
$$
$$ \cdot \lambda_{N_i}\, \vert \, N_1,\ldots
,N_{i-1},N_{i+1},\ldots,N_s\rangle,
  \eqno(3.4)
$$
}}
where
\vspace{-14mm}\\
{\h {
$$ \lambda(N_1,\ldots ,N_s;\,N_i) = \sum_{N<N_i}\lambda_N
$$
}}
and \ {\mbox {\h $ \lambda_N$}} are defined by state (3.3).

 Let us give the scalar product on the fermion states space
 according to formula
{\h {
$$
\langle M,\lambda_M \vert \, N,\lambda_N\rangle=
\delta_{MN}\,\delta_{\lambda_M\,\lambda_N}      \eqno(3.5)
$$
}}

Now  the  operators  conjugated  to  \  {\mbox  {\h  $  a_N$}}   \   and
 their commutational properties
 are found by the standard way by Eq.(3.3) - (3.5).
We have:
  {\h {
  $$  [a^+_M,\,a_N]=\delta_{MN}, \qquad [a_M,\,a_N]=0 \eqno(3.6)
  $$
}}

  Further we shall represent the operators  \ {\mbox {\h $ \{a^+_N,a_N\}$}} \ ,
  Fermi-states and their scale product as in the Appendix.

\baselineskip=18pt
  Let us make the following Hypothesis , which is the base for a further
  development of the dynamic quantization method: \\
 {\underbar {Hypothesis A}}
  {\it { The complete set of one-fermion states
  \ {\mbox {\h $ \vert \, N,1\rangle $}} \ can be chosen so that}}
{\h {
$$
  \langle M,1 \vert \, \int\,d^2x \bar {\psi}(x)\,\Gamma^0(x)\,\psi(x)
  \vert \, N,1 \rangle =\delta_{MN}                        \eqno(3.7)
$$
}}
$$
$$

Emphasize that here and further in this Section averaging is made only
in fermion fluctuations and the fermion functional measure is given in (A4).

  Explain the naturallity of the Hypothesis A.

  Consider the unitary transformation
\vspace{-12mm}\\
{\h {
$$
 \vert \, N, 1\, \rangle ^{\prime} =
  \sum_M S_{NM}\, \vert \, M, 1\, \rangle, \qquad  { }^{\prime}
  \langle N, 1 \,\vert=
  \sum_M  \langle M, 1 \,\vert \,S^+_{NM},
$$
$$   \sum_M S^+_{MN} S_{ML}=\delta_{NL},  \eqno(3.8)
$$
}}
\vspace{-3mm}\\
which conserves the form
 \ {\mbox {\h $ \sum_N \langle N, 1 \,\vert \, N, 1\, \rangle $}} .
 This transformation
 conserves also all above formulae of this Section.
With the help of the unitary transformation  (3.8) one can reach that the
hermithian matrix
{\h {
$$
  \langle M,1 \vert \, \int\,d^2x \,\bar {\psi}(x)\,\Gamma^0(x)\,\psi(x)\,
  \vert \, N,1 \rangle                         \eqno(3.9)
$$
}}
\vspace{-3mm}\\
be diagonal. Note that matrix \ (3.9) \ one can always regard as
positively de-\\ fined.
\footnote{ \large {
\baselineskip=14pt
The necessity of this condition is clear from CR (2.19) for fermion fields.
Indeed denote by $\vert\,\Lambda\rangle$ \ and \ $u(x)$ \
 the arbitrary fermion state and Bose spinor field respectively, and
 \ $\vert\,\Lambda_-,u\rangle = \int d^2x\,(u^+\,\psi)\vert\,\Lambda \rangle,
\ \vert \Lambda_+,u\rangle = \int d^2x\,(\psi^+\,u)\vert\,\Lambda\rangle$ \ .
Then from CR (2.19) we have: \ $\langle\Lambda \vert\, \int d^2x\,u^+\,
(\gamma^0 \Gamma^0)^{-1}\,u\vert\, \Lambda \rangle = \langle\Lambda_+,u
\vert\,\Lambda_+u\rangle + \langle\Lambda_-,u\vert\,\Lambda_-,u\rangle>0$ \ .
Thus the inclusion the matter in gravity leads to serious restrictions
for quantum fluctuations of fields having geometrical sense.
In particular the condition
$det g_{ij}>0$ arises.}}
\baselineskip=18pt
Then taking into account Eq.(3.5) and the form of fermion measure (A4),
 we come to formula (3.7) of Hypothesis A.

                  It follows immediately From Hypothesis A that the fields
 \ {\mbox {\h $ \psi_N(x)$}} \ are linearly independent.

\baselineskip=18pt
  Now we expand the fermion fields:
\vspace{-15mm}\\
 {\h {
 $$ \psi(x)=\sum_N \psi_N(x)\,a_N, \qquad
   \bar{\psi}(x)=\sum_N a^+_N\,\bar{\psi}_N(x)      \eqno(3.10)
 $$
 }}
\vspace{-5mm}\\
  It is not difficult to show that the following CR take place:
{\h {
  $$ [a_N,\,\psi_M]=[a^+_N,\,\psi_M]=0        \eqno(3.11)
  $$
  }}

  Indeed, from the definition of the field \ {\mbox {\h $ \psi_M$}} \
  it follows that
  \ {\mbox {\h $ \psi_M(x)$}} \ can contain operators \ {\mbox {\h $ a^+_L$}}
\
  and \ {\mbox {\h $ a_N$}} \ only in identical degrees. We think that
  the operators \ {\mbox {\h $ a^+_L$}} \ and \ {\mbox {\h $ a_N$}} \
  containing in the fields
  \ {\mbox {\h $ \psi_M(x)$}} \ are normally arranged. It is easy to see
  that in fact the field
  \ {\mbox {\h $ \psi_M(x)$}} \ does not contain the operator
  \ {\mbox {\h $ a_N$}} \ .
 Otherwise Eq.(3.1) and (3.10) should mean that
{\h {
$$
  \psi_N(x)=\psi_N(x)+\sum_M \frac{\partial \psi_M(x)}{\partial a_N}\,a_M ,
$$
}}
But this is impossible in consequence of linearly independence of the fields
 \ {\mbox {\h $ \psi_M$}} \ .

  One can rewrite relation (3.7) with (3.11) in the following form:
 {\h {
 $$
 \int\,d^2x\,\bar{\psi}_M\,\Gamma^0 \psi_N=\delta_{MN}   \eqno(3.12)
 $$
}}
    Since the set of fields \ {\mbox {\h $ \{\psi_M(x)\}$}}  \  is
complete
    relations (3.12) are equivalent to the following equality:
\vspace{-14mm}\\
{\h {
 $$ \sum_N \psi_N(x)\,\bar{\psi}_N(y)=
 \delta^{(2)}(x-y)\,(\Gamma^0)^{-1}(y)         \eqno(3.13)
 $$
 }}
Further
\vspace{-14mm}\\
{\h {
$$
  [\psi (x),\,{\bar {\psi}}(y)]=\delta^{(2)}
(x-y)(\Gamma^0)^{-1}(y)=
$$
\vspace{-9mm}\\
$$
  =\sum_{MN}[\bar {\psi}_M(y),\psi_N(x)]\,a^+_M\,a_N+\sum_N \psi_N(x)\,
  \bar {\psi}_N(y)                               \eqno(3.14)
$$
\vspace{-1mm}\\
}}
 The first from equalities (3.14) is taken from (2.17), the second one
 is obtained by Eqs.(3.6) and (3.10). Comparing relations
 (3.13) and (3.14) we find the CR:
\vspace{-11mm}\\
 {\h {
 $$
 [\bar {\psi}_M(y),\psi_N(x)]=0                  \eqno(3.15)
 $$
 }}
 Analogously we get:
 {\h {
 $$
 [\psi_M(y),\psi_N(x)]=0                  \eqno(3.15')
 $$
 }}

\baselineskip=18pt
     Now reinforce the Hypothesis A, completing it by the following:

{\underbar {Hypothesis B}}
{\it { The set of states \ {\mbox {\h $ \vert \, N,\lambda_N \rangle$}} \
 is gauge invariant, that is}}
  {\h {
  $$ [a_N,\,\chi_a]=0, \qquad [a_N,\,\phi_a]=0          \eqno(3.16)
  $$
\vspace{-3mm}\\
  }}
\centerline{ }

  Note that if any formulae are written out for \ {\mbox {\h $a^+_N$}} \
 or  \ {\mbox {\h $\bar{\psi}$}} \ then the analogous formulae for
  \ {\mbox {\h $a_N$}} \ or  \ {\mbox {\h $\psi$}} \ are obtained by the
  Hermithian
 conjugation.

  Comparison of Eqs.(3.16) with the analogous  equations in usual gauge theory
(see Eqs.(3.17) in [1]) shows that the application of the dynamic quantization
to
general covariant theories is the most natural. This follows  from  (i)
in general
covariant theories all dynamics is reduced to gauge transformations; (ii)
the key point of the dynamic quantization method is the selection of the
gauge invariant operators like the annihilation and
creation operators introduced here. As is shown in [1]
 in the usual gauge theory the gauge
invariant operators playing the same role in dynamic quantization have the
phase factor depending on time significant  near  cutoff
impulse.

\baselineskip=18pt
  It \ follows \ from \ equations \ of motion (see Section 5) \ that
the \ operator \\
\ {\mbox {\h $ \int d^2x\,\bar {\psi}\,\Gamma^0 \psi$}} \
is conserved. Thus, Hypothesis B results that all relations of this Section
are conserved in time. Moreover, the fields
\ {\mbox {\h $ \psi_M(x)$}} \ satisfy the same equation as the field
 \ {\mbox {\h $ \psi(x)$}}:
{\h {
 $$
 i\,\Gamma^{\mu}\,\nabla_{\mu}\,\psi_M=\frac{i}{2}\,(8\pi G)\,
 \varepsilon_a{ }^{bc}\,e_c\,\gamma^a\,\psi_M \,\chi_b         \eqno(3.17)
 $$
}}

\baselineskip=18pt
 Using expansion (3.10) and the completeness property (3.12) one can
 express the annihilation and creation operators in the form:
 {\h {
 $$  a^+_N=\int\,d^2x\,\bar{\psi}\,\Gamma^0\psi_N    \eqno(3.18)
 $$
\vspace{-3mm}\\
 }}

\centerline{ }
\centerline{\it { 3.2 The Imposition of regularizing Constraints.}}

\centerline{ }

\baselineskip=18pt
      As Witten showed [5]  ,  if  the  matter  is  absent,  the
quantum  gravity in 3-dimensional  space-time  is  not  only
renormalizable but also finite theory
(in the framework of the Feynman theory). However, the coupling with
the Dirac fields makes the Feynman quantization impossible, since in the Witten
variables \ {\mbox {\h $({\cal P}^i_a,\,\omega^a_i)$}} \ the
zeroth approximation
for the Dirac field is absent. The reason is that according to Witten
the variables \ {\mbox {\h ${\cal P}^i_a$}} \ and
\ {\mbox {\h $\omega^a_i$}} \ fluctuate nearby classic values
\ {\mbox {\h ${\cal P}^i_a=0$}} \ and  \ {\mbox {\h $\omega^a_i=0$}} \  .
This point of view is also accepted in our work. But from (2.14) it is seen
that in this case the fermion contribution to the Hamiltonian contains only
the terms higher than quadratic order in the fundamental fields.
Therefore, the Fermi part of Hamiltonian can be accounted only as a whole by PT
what is impossible in the Feynman theory.

  Another picture is under the dynamic quantization. The dynamic quantization
  can retain any finite number of fermion degrees of freedom in the theory.
 At the same time the rest of fermion degrees of freedom are removed
completely
 from dynamics. Therefore, the theory becomes regularized.
\baselineskip=18pt
   Moreover, since the number of the retained degrees of freedom can be "enough
   small", so the development of PT by expansion in the fermion part of
   the Hamiltonian becomes possible.

    For example, let the surface \ {\mbox {\h $x^0=Const$}} \ be
  a compact Riemann surface of the genus \ {\mbox {\h $g$}} \ ,
  which we denote by \ {\mbox {\h $\Sigma$}} \ .
   Since \ {\mbox {\h $\Sigma$}} \ is a compact surface so one can think the
set
   of indices \ {\mbox {\h $N$}} \  coincides with the set \ {\mbox {\h $Z$}} \
 .

       The regularization of the theory is achieved by imposing of the
       infinite set of the second class constraints:
 \vspace{7mm}\\
    \ \ \ \ {\mbox {\h $a_N^+= 0, \ \ a_N=0$}} \  \  at
    \ \ {\mbox {\h $|\,N\,|>N_0 \in Z \qquad  \qquad (3.19)$}} \
 \vspace{7mm}\\
  Constraints (3.19) mean that the states
 \ {\mbox {\h $\vert \, N,1\rangle$}} \ with
 \ {\mbox {\h $|\,N\,|>N_0$}} \ are dropped.

   We perform the regularization of any operator \ {\mbox {\h $A$}} \
   by the following order: at first we arrange normally the fermion operators
    \ {\mbox {\h $a_N^+$}} \ and \ {\mbox {\h $a_N$}} \ containing in the
    operator \ {\mbox {\h $A$}} \  and then impose constraints (3.19).
    The resultant operator is designated as \ {\mbox {\h $A_{reg}$}}.

    The concrete choice of the fields \ {\mbox {\h $\{\psi_N\}$}} \
    at \ {\mbox {\h $|\,N\,|<N_0$}} \ and also the method of "vacuum filling"
    (see below) means the choice of the initial physical conditions.

\baselineskip=18pt
}

\centerline { }
\centerline { }
\centerline {{\bf 4. The Dirac Commutational Relations.}}
\centerline { }

{\large {
\baselineskip=18pt
    Under the second class constraints (3.19) the classical CR (2.17) are
    to be replaced by the corresponding Dirac CR. Here this problem is solved.

   Denote by the symbol \ {\mbox {\h ${\cal G}$}} \ the Grassmann algebra
   with generators (A1). The elements of the algebra
   \ {\mbox {\h ${\cal G}$}} \ are fermion states labelled by
    large Greek letters \ {\mbox {\h $\Lambda,
      \Sigma,\Pi,\ldots$}} \ .
      For any operator
 \ {\mbox {\h $A$}} \ denote
  {\h {
  $$
  \langle \Lambda\vert \, A\vert \, \Sigma \rangle =A_{\Lambda\Sigma}
  $$
  }}
  Once more pay attention to the fact that here the averaging is performed
  only in fermion fluctuations according to rule (A3).

  By the definition in regularised theory
  the space of fermion states
  \ {\mbox {\h ${\cal G}^{\prime}$}} \  is the subalgebra of algebra
  \ {\mbox {\h ${\cal G}$}} \  with generators taken from set (A1),
  satisfying the condition \ {\mbox {\h $|\, N \,|<N_0$}} \ .
  The elements of the space  \ {\mbox {\h ${\cal G}^{\prime}$}} \
  are numbered by the primed Greek letters
 \ {\mbox {\h $\Lambda^{\prime},\Sigma^{\prime},\Pi^{\prime},\ldots$}} \ .

  Note that in our case
 the space \ {\mbox {\h ${\cal G}^{\prime}$}} \  is finite-dimensional due to
 the compactness of space surface.

    According to the definition of the regularized operators (see Section 3.2)
\vspace{-13mm}\\
{\h {
$$
   (A_{reg})_{\Lambda^{\prime},\Sigma^{\prime}} =
    A_{\Lambda^{\prime},\Sigma^{\prime}}       \eqno(4.1)
$$
}}
\vspace{-3mm}\\
  In a regularized theory we shall omit the index $reg$ in
  designation of regularized operators.

  The Dirac CR corresponding to the constraints (3.19) are defined
  according to
\vspace{-13mm}\\
{\h {
$$
  [A,\,B]^*_{\Lambda^{ \prime} \Sigma^{ \prime}}=
  \sum_{\Pi^{ \prime}}(A_{\Lambda^{ \prime} \Pi^{ \prime}}B_{\Pi^{
\prime}\Sigma^{ \prime}}-
  B_{\Lambda^{ \prime} \Pi^{ \prime}}A_{\Pi^{ \prime}\Sigma^{ \prime}})
\eqno(4.2)
$$
}}
\vspace{-5mm}\\
\baselineskip=18pt
  Note that as a consequence of Hypothesis B we have for Hamiltonian
of the system:
\vspace{-17mm}\\
  {\h {
  $$
    H_{\Lambda\Sigma}=\delta_{\Lambda\Sigma}\,H_{\Sigma\Sigma}
$$
}}
\vspace{-5mm}\\
  From here and definition (4.2)
 for any
  regularized operator \ {\mbox {\h $A$}} \ there is the following important
  equation:
\vspace{-13mm}\\
  {\h {
  $$
  [A,\,H]^*_{\Lambda^{\prime} \Sigma^{\prime}}=
  [A,\,H]_{\Lambda^{\prime} \Sigma^{\prime}}   \eqno(4.3)
  $$
  }}
\vspace{-4mm}\\
  Here the right hand side is obtained formally by classic
  CR (2.19). Eq.(4.3) shows that the regularized Heisenberg equations for the
  fields \ {\mbox {\h $\omega_i^a, \, {\cal P}_a^i$}} \
and \ {\mbox {\h $\bar{\psi}, \, \psi$}} \   coincide
(modulo operator permutations)
 with the formal one or classic equations and regularized
algebra of gauge transformations coincide with the corresponding formal
algebra.

   Using definition (4.2) and the results of Section 3 we obtain:
\vspace{-14mm}\\
{\h {
$$  [\psi(x),\,\bar{\psi}(y)]^*
   =\sum_{|N|<N_0}\,\psi_N(x)\,\bar{\psi}_N(y)     \eqno(4.4)
    $$
\vspace{-3mm}\\
    }}

    We state that equations (4.3) and (4.4) define the Dirac CR for
    all fundamental fields. Let us show that.

    For reduction introduce the following designations:
{\h {
$$ \widehat{\nabla}_i\,{\cal P}_a^i=\partial_i\,{\cal P}_a^i-
 \frac{1}{2}\,\varepsilon_{ab}{ }^c(\omega_i^b\,{\cal P}_c^i+
 {\cal P}_c^i\,\omega_i^b),
 $$
\vspace{-1mm}
 $$ \chi_a=\widehat{\nabla}_i\,{\cal P}_a^i+T_a,
 $$
\vspace{-1mm}
 $$  F^a=\frac{1}{2}\,\varepsilon_{ij}(\,\partial_i\omega_j^a-
  \partial_j\omega_i^a-\varepsilon^a{ }_{bc}\omega_i^b\omega_j^c) ,
  $$
\vspace{-1mm}
$$ \phi^a=F^a+R^a        \eqno(4.5)
$$
\vspace{-3mm}\\
}}
 Thus, the quantities \ {\mbox {\h $T_a$}} \ and \ {\mbox {\h $R_a$}} \
are the fermion parts of operators \ {\mbox {\h $\chi_a$}} \ and
\ {\mbox {\h $\phi_a$}}, \ respectively.

\baselineskip=18pt
   Let us search the rest of the Dirac CR by expansion in constant
  \ {\mbox {\h $G$}} using Eqs.(4.3) and (4.4). The expansion in
 \ {\mbox {\h $G$}} \ is equivalent to the expansion in the quantities
\ {\mbox {\h $R^a$}} \ in (4.5), which are the fermionic part of the
constraints \ {\mbox {\h $\phi_a$}}.
 Let \ {\mbox {\h $\lambda_F$}} \ be a length
 of the order of the minimum wave length of the functions \ {\mbox {\h
$\psi_N$}} \
 for
 \ {\mbox {\h $|N|<N_0$}} .
Since in our theory the number of
fermion degrees of freedom is limited, this expansion is mathematically
correct under a significant smallness of the dimensionless parameter
\ {\mbox {\h $G\,\lambda_F^{-1}$}} \ . This expansion is used in
Section 6 to solve the equations of motions.

\baselineskip=18pt
  To begin the expansion in constant \ {\mbox {\h $G$}} \ ,
it is necessary to arrange by parameter \ {\mbox {\h $G$}} \
the following quantities:
\vspace{-14mm}\\
{\h {
$$ \psi, \bar{\psi}\sim 1, \ \ \omega_i^a \sim 1, \ \ {\cal P}_a^i \sim
  (G)^{-1},
  $$
\vspace{-3mm}
$$ [\psi,\,\bar{\psi}]^*\sim 1, \ \ [{\cal P}_a^i,\,{\cal P}_b^j]^*\sim 1,
\ \ \ [\omega_i^a,\,\omega_j^b]^*\sim G^2,
 $$
\vspace{-3mm}
 $$
  \ \ [\psi,\,{\cal P}_a^i]^* \sim 1, \ \ [\psi,\,\omega_i^a]^*\sim G,
\ \ [{\cal P}_a^i,\,\omega_j^b]^*\sim 1    \eqno(4.6)
$$
\vspace{-3mm}\\
}}
The commutator \ {\mbox {\h $[\psi,\,\bar{\psi}]^*$}} \ is known strictly
according to (4.4).

\baselineskip=18pt
 From equation

\baselineskip=18pt
\large {
\baselineskip=18pt

{\h {
$$ [\psi(x),\,\chi_a(y)]^*=[\psi(x),\,\chi_a(y)]
$$
}}
(see. (4.3)) we extract the contribution of the zeroth order in \ {\mbox {\h
$G$}}
{}.
For this we must neglect the quantities
\vspace{-14mm}\\
{\h {
$$ [\psi,\,\omega_i^a]^*\sim G, \ \ [\psi,\,\sqrt{g}\,e_a^0]^*\sim G
$$
\vspace{-3mm}\\
}}
(see (2.15) and (4.6)). We find:
{\h {
$$ \nabla_i([\psi(x),\,{\cal P}_a^i(y)]^*)^{(0)}=
$$
\vspace{-3mm}
$$
 \frac{1}{2}\,\sum_{|N|>N_0}\psi_N(x)\,(\,\bar{\psi}_N
\sqrt{g}e_a^0\psi)\,(y)                              \eqno(4.7)
$$
\vspace{-1mm}\\
}}
Here superscripts \ {\mbox {\h $(0), (1),\ldots$}} \  denote the order of
the quantity in the parameter \ {\mbox {\h $G$}} .

   From the equation
{\h {
$$ [\psi(x),\,\phi_a(y)]^*=[\psi(x),\,\phi_a(y)]
$$
}}
we extract the contribution of the order \ {\mbox {\h $G$}} .
Taking (4.6), (2.14c) and (2.17) into account, we obtain
\vspace{-14mm}\\
{\h {
$$ \varepsilon_{ij}\,\nabla_i(y)\bigl([\psi(x),\,\omega_j^a(y)]^*\bigr)^{(1)}=
$$
\vspace{-3mm}
$$ =\varepsilon_{ij}\,\nabla_i(y)\,[\psi(x),\,\omega_j^a(y)]+
   \sum_{|N|>N_0}\psi_N(x)\,R^a_{\bar {N}}(y)                     \eqno(4.8)
$$
\vspace{-1mm}\\
}}
     Here the quantity \ {\mbox {\h $R^a_{\bar {N}}(y)$}} \  ( or
 \ {\mbox {\h $R^a_N(y)$}} \   ) is obtained from  \ {\mbox {\h $R^a(y)$}} \
 by the substitution of \ {\mbox {\h ${\bar {\psi}_N(y)}$}} \ for
  \ {\mbox {\h ${\bar {\psi}(y)}$}} \  ( or  \ {\mbox {\h $\psi_N(y)$}} \
 for  \ {\mbox {\h $\psi(y)$}}) .

  Using Eq.(4.3) for \ {\mbox {\h $A=
 \widehat{\nabla}_i\,{\cal P}_a^i$}} \ we find
\vspace{-14mm}\\
{\h {
$$  \nabla_i(x)\,\nabla_j(y)\,
\bigl(\, [{\cal P}_a^i(x),\,{\cal P}_b^j(y)]^*\bigr)^{(0)}=
$$
\vspace{-3mm}
$$ =-\frac{1}{4}\,\sum_{|N|>N_0}\{(\bar{\psi}\sqrt{g}\,e_a^0\psi_N)
(x)\,(\bar{\psi}_N\sqrt{g}\,e_b^0\psi)(y)-
$$
\vspace{-3mm}
$$  -(\bar{\psi}\sqrt{g}\,e_b^0\psi_N)(y)\,(
\bar{\psi}_N\sqrt{g}\,e_a^0\psi)(x)\} \ ,    \eqno(4.9)
$$
}}
{\h {
$$ \varepsilon_{jk}\,\nabla_i(x)\,\nabla_j(y)\,
 \bigl(\, [{\cal P}_a^i(x),\,\omega^b_k(y)]^*\bier)^{(1)}=
 $$
\vspace{-3mm}
 $$ \frac{1}{2}
 \,\sum_{|N|>N_0}\{R^b_N(y)\,(\bar{\psi}_N\,\sqrt{g}\,e_a^0\psi)(x)-
 (\bar{\psi}\sqrt{g}\,e_a^0\psi_N)(x)\, R^b_{\bar {N}}(y)\}   \eqno(4.10)
$$
\vspace{-6mm}\\
}}

     By this way we obtain expressions for all Dirac CR in the lowest order
     in the parameter \ {\mbox {\h $G\,\lambda_F^{-1}$}} . One can then develop
     the iterations in
 \ {\mbox {\h $G\,\lambda_F^{-1}$}} \ , using (4.3) and the initial values
 for the
 Dirac commutators obtained here.

   The defined Dirac CR possesses all the necessary properties (2.18),
   (2.19). It follows from definition (4.2).
\baselineskip=18pt

\centerline{ }
\centerline{ }
\centerline {\bf 5. Heisenberg Equations and the State Vectors.}
\centerline{ }

\baselineskip=18pt
Equations (4.3) show that the regularized Heisenberg equations
can be obtained using formal CR (2.17).

Here two questions arise:\\
1). How does one define a product of
operator fields at one spatial point \ {\mbox {\h $x$}} \ ? \\
2). How does one order the operator fields at one spatial point
\ {\mbox {\h $x$}} \ in the
Heisenberg equations obtained, since the CR (2.17) differs from the Dirac CR?

We have the following answer to the first question.

In consequence of the constraints (3.19) the Dirac fields
\ {\mbox {\h $\psi$}} \ and
 \ {\mbox {\h $\bar{\psi}$}} \  are smooth and thus, their product
 ( in any order and probably, at the same point
 \ {\mbox {\h $x$}} \ ) is regular.

\baselineskip=18pt
If in the theory under consideration the fermion degrees of freedom
were absent, there would be no local degrees of freedom.
In this case there would be only global degrees of freedom,
connected with fundamental group of spatial surface.
If the spatial surface is a  compact Riemann surface of
genus  \ {\mbox {\h $g$}}  \
its  fundamental group
has  \ {\mbox {\h $2g$}} \  generators
  \ {\mbox {\h $a_i, b_j, i,j=1,\cdots ,g$}} \ with the only
one constraint
 {\h {
$$
a_1\,b_1\,a_1^{-1}b^{-1}_1\cdots a_g\,b_g\,a_g^{-1}b_g^{-1}=1    \eqno(5.1)
$$
 }}
Let's introduce the notations
{\h {
$$ U_i\,(V_i)=\widehat{T}\,\exp\frac{i}{2}\,\oint_{\gamma_i(\delta_i)}\,
  \omega_{ai}\gamma^a\,dx^i\subset SO(2,1)       \eqno(5.2)
$$
}}
where  \ {\mbox {\h $\gamma_i(\delta_i)$}} \ is any representative
of the class  \ {\mbox {\h $a_i(b_i)$}} \  and
  \ {\mbox {\h $\widehat{T}$}} \ denotes the ordering operator along
the integration path. Since, in the absence of fermions,
  \ {\mbox {\h $F_a=0$}} \ quantities
(5.2) do not depend on the representatives of the classes  \ {\mbox {\h
$a_i(b_i)$}} \ .
The quantities (5.2) satisfy Eq.(5.1):
{\h {
$$ U_1V_1U_1^{-1}V_1^{-1}\cdots U_gV_gU_g^{-1}V_g^{-1}=1  \eqno(5.3)
$$
}}
Eq.(5.3) are the single constraints for quantities (5.2) [5].
                                 The wave functions of the system
  \ {\mbox {\h $\Psi(U_i,V_j)$}} \ in the absence of fermion
  fields depend only on
quantities (5.2); the   \ {\mbox {\h $ \Psi$}} \ are defined on
the hypersurface (5.3)
and are invariant relative to the transformations (5.4).
{\h {
$$U_i\longrightarrow E^{-1}U_iE, \ \ \ V_j\longrightarrow E^{-1}V_jE
\eqno(5.4)
$$
}}
where \ {\mbox {\h $E$}} \ is a certain element from the group
 \ {\mbox {\h $SO(2,1)$}} .
There are also constraints:
\vspace{-14mm}\\
{\h {
$$ \widehat{\nabla}_i\,{\cal P}^i_a\,\Psi=0, \ \qquad \
      F_a\,\Psi=0                                  \eqno(5.5)
      $$
\vspace{-3mm}\\
      }}
It follows from above that in the absence of fermion fields
the small-scale fluctuations of fields \ {\mbox {\h $\omega^a_i$}} \ and
 \ {\mbox {\h ${\cal P}^i_a$}} \ do not play any role, they can be removed.
Under small-scale fluctuations we mean the fluctuations with the wavelength
 \ {\mbox {\h $\sim \lambda$}} \  for which
 {\h {
$$ \lambda<\lambda_0\ll s_i,\,r_i \qquad i=1,\cdots ,g     \eqno(5.6)
$$
}}
\baselineskip=18pt
where \ {\mbox {\h $s_i, r_i$}} \ are the characteristic length of cycles
  \ {\mbox {\h $\gamma_i,\,\delta_i$}} \ in some metrics.
  The "removing" of the small-scale
 fluctuations of the fields \ {\mbox {\h $\omega^a_i$}} \ and
 \ {\mbox {\h ${\cal P}^i_a$}} \ implies
 that in the corresponding commutator in (2.17) the
 \ {\mbox {\h $\delta$}}-function is replaced by a smoothed
  \ {\mbox {\h $\delta$}}-function , which
we designate as \ {\mbox {\h $\delta_{\lambda_0}$}}  . The
concrete method of smoothing
of \ {\mbox {\h $\delta$}}- function is of no importance here.
It is important only that
  \ {\mbox {\h $\delta_{\lambda_0}$}} \ acts on smooth functions
  ( which are change
  noticeable on the scales much greater than \ {\mbox {\h $\lambda_0$}} \ )
  in the same way
  as the \ {\mbox {\h $\delta$}}-function.

\baselineskip=18pt
 In our case, in the presence of the fermion degrees of freedom,
 one must proceed analogously. Condition (5.6) should be
 supplemented by
 {\h {
$$  \lambda_0 \ll \lambda_F     \eqno(5.7)
$$
}}
\baselineskip=18pt
after which the regularization scheme for fields \ {\mbox {\h $\omega^a_i$}} \
and
 \ {\mbox {\h ${\cal P}^i_a$}} \  remains valid. The quantity
\ {\mbox {\h $\lambda_F$}} \ is defined in the previous
 Section. Therefore the answer to first question
 is given.

\baselineskip=18pt
 The regarded regularization of the Bose degrees of freedom is in agreement
 with PT developed here. Indeed, in the zeroth order (see (6.3) - (6.9)) the
 solutions of the Heisenberg equations are linearly expressed through the
initial
 values of the fields. This allows are to remove easily
 the small-scale fluctuations
 of the fields \ {\mbox {\h $\omega^a_i$}} \ and \ {\mbox {\h ${\cal P}^i_a$}}
\ .
 Since in our PT the
 expansion is performed by the whole fermion part of the Hamiltonian, then in
 consequence with condition (5.7) the PT conserves the sense of the
  regularization of the Bose degrees of freedom also in higher orders of
 expansion.

\baselineskip=18pt
The answer to the second question will be given after some formal calculations
in anregularized theory.

 We say that a given system is formally quantized if the formal
 algebra of the operators
 \ {\mbox {\h $(\chi_a,\,\phi_a)$}} \ is closed and the
 structure functions are placed
all to the left (or right) of the generators  \ {\mbox {\h
$(\chi_a,\,\phi_a)$}} \
as in the following equation
{\h {
 $$
 [\phi_a(x),\,\phi_b(y)]=\delta^{(2)}(x-y)\cdot
 $$
\vspace{-3mm}
 $$\cdot \{\,f_{ab}{ }^c(x)\,\phi_c
 (x)+g_{ab}{ }^c(x)\,\chi_c(x)\,\}
$$
\vspace{-8mm}\\
 }}

\baselineskip=18pt
 We shall carry out formal calculations of the necessary commutators,
 using the relations of the form
\vspace{-14mm}\\
 {\h {
$$
 \psi_{\sigma}(x)\,\bar {\psi}_{\rho}(x)=-\bar{\psi}_{\rho}(x)\,\psi_{\sigma}
 (x)+\delta^{(2)}(0)\,(\Gamma^0)^{-1}_{\sigma \rho}(x),  $$
 $$
 \omega^a_i(x)\,{\cal P}^j_b(x)={\cal P}^j_b(x)\,\omega^a_i(x)+
 \delta^{(2)}(0)\,i\,\delta^a_b\,\delta^j_i                        \eqno(5.8)
 $$
 }}
 e.t.s., which follow formally from the CR (2.17). This approach makes it
possible to
 calculate the (operator) coefficient of the symbol \ {\mbox {\h
$\delta^{(2)}(0)$}} \
 arising as a result of permutations of the field operators in various
 expressions. In this way we arrive at the conclusion that system (2.14)
  is formally quantized only for the value
  {\h {
$$     s=\frac{5}{8}                     \eqno(5.9)
$$
}}
\baselineskip=18pt
 Thus, the requirement of the formal quantizability of the system  fixes
uniquely the ordering
 of the operator fields in the Hamiltonian.

\baselineskip=18pt
 The formulae given below are valid for \ {\mbox {\h $s=\frac{5}{8}$}} .

\centerline{ }
\centerline{ \it 5.1 The Equations of Motion and
the Gauge Transformations Algebra.}
\centerline{ }

\baselineskip=18pt
 The formal calculations are carried out in stages. In the first stage
 we find the Heisenberg equations \ {\mbox {\h $i\,\dot{A}=[A,\,H]$}} \
 for fundamental fields:
 {\h {
 $$
 i\,\Gamma^{\mu}\,\nabla_{\mu}\,\psi=\frac{i}{2}\,(8\pi G)\,
 \varepsilon_a{ }^{bc}\,e_c\,\gamma^a\,\psi \,\chi_b ,
 $$
 $$
 i \nabla_{\mu}\bar{\psi}\,\Gamma^{\mu}=\frac{i}{2}\,
 (8\pi G)\,\varepsilon_a{ }^{bc}\,e_c\,\chi_b\,{\bar{\psi}}\,\gamma^a,
$$
 $$
 \dot {\cal P}^i_a-\varepsilon_{ab}{ }^c\,\omega^b\,{\cal P}^i_c+
 \frac{1}{8\pi G}\,\varepsilon_{ij}\,(\partial_je_a-\varepsilon_{ab}{ }^c\,
 \omega^b_j\,e_c)+
 $$
 $$
 +\frac{1}{2}\,(8\pi G)\,\varepsilon_a{ }^{bc}\,e_c\,
(\bar{\psi}\,{\cal P}^i_b\,\psi)=0,
$$
 $$
 i\,{\dot {\omega}}^a_i=i\,(\partial _i\omega^a-\varepsilon^a{ }_{bc}\,
 \omega^b_i\,\omega^c)+
 $$
 $$
 +\frac{1}{2}\,(8\pi G)\,\varepsilon_b{ }^{ca}\,e_c\,(\bar{\psi}\,\gamma^b
 \nabla_i\,\psi-\nabla_i\bar{\psi}\,\gamma^b\,\psi)+  $$
 $$
 +\frac{i}{2}\,(8\pi G)^2\,\varepsilon_b{ }^{ca}\,
 \varepsilon_{ij}\,(
\bar{\psi}\,{\cal P}^j_c\,\gamma^b\,[\psi,\,H_{\phi}]-
$$
$$
-[\bar{\psi},\,H_{\phi}\,]\,\gamma^b\,{\cal P}^j_c\,\psi)       \eqno(5.10)
 $$
 }}

 In the classic limit all the Equations (5.10) with the exception of the
 Dirac equations coincide with Euler-Lagrange equations  obtained by
 the variations of the action (2.7). Equations (5.10) for the Dirac fields
 differ inessentially from the Dirac equations, by only a term that is equal
 to zero in a weak sense.

\baselineskip=18pt
 Now we can show that the quantity
  \ {\mbox {\h $\int\,d^2x \,\bar{\psi}\,\Gamma^0\psi$}} \ is conserved.
  With the help of Eq.(5.10) we obtain:
\vspace{-14mm}\\
  {\h {
  $$
   \partial_{\mu}\,\Gamma^{\mu}-
  \frac{i}{2}\,\omega_{a\mu}\,\gamma^a\,\Gamma^{\mu}+
  \frac{i}{2}\,\Gamma^{\mu}\,\gamma^a\,\omega_{a\mu}=
   (8\pi G)\,\varepsilon_b{ }^{ca}\,e_a\,\gamma^b\,\chi_c
  $$
\vspace{-3mm}\\
  }}
  From here and the two first equations of (5.10) we find that
  {\h {
  $$
  \frac{d}{dt}\,\int d^2x\,\bar{\psi}\,\Gamma^0\,\psi=0
  $$
  }}
  According to Eq.(4.3) the last equation is valid both in formal and
  regularized theories.

\baselineskip=18pt
 Let \ {\mbox {\h $A$}} \ be any fundamental field.
 We have the Jacobi identity:
 {\h {
$$ [\,[\phi_a(x),\,\phi_b(y)],\,A]=
$$
$$
[\phi_a(x),\,[\phi_b(y),\,A]\,]-
[\phi_b(y),\,[\phi_a(x),\,A]\,]                \eqno(5.11)
$$
\vspace{-3mm}\\
}}
\baselineskip=18pt
 The right-hand side in (5.11) is calculated by Eq.(5.10).
 Knowing the right-hand side of the expression (5.11), and also the classic
 Poisson bracket \ {\mbox {\h $[\phi_a(x),\,\phi_b(y)]$}} \ , as
 a  result  of  time
 consuming
 calculations we can establish the quantum formal commutator
 \ {\mbox {\h $[\phi_a(x),\,\phi_b(y)]$}} . Here we use in all the calculations
only
 Eqs.(5.10) and formal algebra (5.8). Note that in all the described formal
 calculations the symbol  \ {\mbox {\h $\delta^{(2)}(0)$}} \ is encountered
 to a power
 no higher than the first, and symbols of the form
   \ {\mbox {\h $\delta^{\prime}(0)$}} \
 are absent in the calculations. Also, the final results do not contain
 the indeterminate symbol  \ {\mbox {\h $\delta^{(2)}(0)$}} .

\baselineskip=18pt
 The answer is
\vspace{-14mm}\\
{\h {
 $$
 [\chi_a(x),\,\chi_b(y)]=-i\,\delta^{(2)}(x-y)\,\varepsilon_{ab}{ }^c\,
 \chi_c(x) \  ,       \eqno(5.12a)
 $$
\vspace{-4mm}
 $$
  [\chi_a(x),\,\phi^b(y)]=-i\,\delta^{(2)}(x-y)\,\varepsilon_{ca}{ }^b\,
 \phi^c(x) \ ,      \eqno(5.12b)
 $$
\vspace{-3mm}
 $$
 [\phi^a(x),\,\phi^b(y)]=\delta^{(2)}(x-y)\,\frac{i}{2}\,(8\pi G)^2\,
 \varepsilon^{ab}{ }_c(\bar{\psi}\,\psi)\,\phi^c+ $$
\vspace{-3mm}
$$
 +\delta^{(2)}(x-y)\,\frac{1}{2}\,(8\pi G)^4\,(\varepsilon^{ab}{ }_d\,
 \varepsilon_g{ }^{hc}-\varepsilon^{abc}\,\varepsilon_{dg}{ }^h)\cdot
$$
\vspace{-3mm}
$$
 \cdot\{\,\bar{\psi}\,{\cal P}^i_c\,\gamma^g\,(\Gamma^0)^{-1}\,\gamma^d\,
 \nabla_i\,\psi-\nabla_i\bar{\psi}\,\gamma^d\,(\Gamma^0)^{-1}
 \gamma^g\,{\cal P}^i_c\,\psi\,\}
 \chi_h +  $$
\vspace{-3mm}
 $$
 +\delta^{(2)}(x-y)\,\frac{1}{4}\,(8\pi
G)^4\,(\varepsilon^{abh}\,\varepsilon_{gd}{ }^f
 -\varepsilon^{ab}{ }_g\,\varepsilon_d{ }^{fh})\cdot   $$
\vspace{-3mm}
 $$
   \cdot \{\,A\,\chi_h\,\bar{\psi}\,\gamma^g\,(\Gamma^0)^{-1}\gamma^d\,
 \psi\,\chi_f+B\,\chi_f\bar{\psi}\,\gamma^g\,(\Gamma^0)^{-1}\gamma^d\,\psi\,
 \chi_h + $$
\vspace{-3mm}
$$
+C\,\bar{\psi}\,\gamma^g\,\chi_h\,(\Gamma^0)^{-1}\gamma^d\,
 \psi\,\chi_f+
  D\,\bar{\psi}\,\gamma^g\,\chi_f\,(\Gamma^0)^{-1}\gamma^d\,
 \psi\,\chi_h +   $$
\vspace{-3mm}
 $$
 +F\,\bar{\psi}\,\gamma^g\,(\Gamma^0)^{-1}\gamma^d\,\chi_h\,\psi\,\chi_f +
  Q\,\bar{\psi}\,\gamma^g\,(\Gamma^0)^{-1}\gamma^d\,\chi_f\,\psi\,\chi_h +
$$
\vspace{-3mm}
$$
 +H\,\bar{\psi}\,\gamma^g\,(\Gamma^0)^{-1}\gamma^d\,\psi\,\chi_h\,\chi_f+ $$
\vspace{-3mm}
 $$
 +(1-A-B-C-D-F-Q-H)\cdot
$$
\vspace{-3mm}
$$
\cdot\bar{\psi}\,\gamma^g\,(\Gamma^0)^{-1}\gamma^d\,
 \psi\,\chi_f\,\chi_h\,\}   \eqno(5.12c)
 $$
\vspace{-3mm}\\
}}
 Here    \ {\mbox {\h $A,\,B,\,C,\,D,\,F,\,Q,\,H$}}  \   are  some  numerical
real
parameters,
 satisfying the following equations:
\vspace{-14mm}\\
 {\h {
 $$
    D+F+2=0,\ \qquad \ \ B+H=1   $$
  $$
    2\,A+D-F=0     $$
  $$
    2\,B+3\,D-F+2\,Q=0       $$
    }}
  All the operators in the right-hand side of Eqs.(5.12) are taken at the
point
   \ {\mbox {\h $x$}} . The right-hand side of (5.12c) is
   formally anti-Hermitian.

\baselineskip=18pt
  The above theory insists that the regularized Heisenberg
  equations for fundamental fields coincide with Eq.(5.10) and the
 values of the commutators
\vspace{-14mm}\\
{\h {
$$ [\chi_a(x),\,\chi_b(y)]^*, \ \ [\chi_a(x),\,\phi^b(y)]^*,
    \ \ [\phi^a_b(x),\,\phi^b(y)]^*
$$
\vspace{-3mm}\\ }}
coincide with the right-hand side of Eqs.(5.12a), (5.12b) and (5.12c),
respectively. Moreover, the right-hand side of Eq.(5.12c) is anti-Hermitian
also in regularized theory.
\baselineskip=18pt

\centerline{ }
\centerline{ \it 5.2 The State Vectors.}
\centerline{ }

\baselineskip=18pt
 By definition, for the wave functions  \ {\mbox {\h $\Psi$}} \  we have
 {\h {
$$ a_N\,\Psi=0, \qquad |N|\,<\,N_0       \eqno(5.13)
$$
}}
  From (5.5) and (5.13) it follows that
  {\h {
$$ \phi_a\,\Psi=0, \qquad \chi_a\,\Psi=0      \eqno(5.14)
$$
}}

    We now have all the means for the construction of the state vectors.
We write out the base state vectors:
{\h {
$$ \Psi_{N_1N_2\ldots N_s}=a^+_{N_1}\,a^+_{N_2}\ldots a^+_{N_s}\,\Psi ,
$$
$$
   |N_i|<\,N_0, \ s=0,1,\ldots                       \eqno(5.15)
$$
}}
 Any state is the superposition of the states (5.15). From relations
 (3.18) and (5.14) it follows immediately that
 {\h {
$$ \chi_a\,\Psi_{N_1N_2\ldots N_s}=0, \ \qquad \phi_a\,\Psi_{N_1\ldots N_s}
 =0 \eqno(5.16)
$$
}}

{\underbar {{\it Remark.}}} Since in the theory under consideration
the problem of the
eigen modes of the Dirac operator has no meaning, it is impossible
to classify the operators   \ {\mbox {\h $a_N$}} \ and
  \ {\mbox {\h $a_N^+$}} \ by their energy.
Therefore, the problem of filling the physical vacuum remains
unsolved here. A criterion permitting one to distinguish the ground and
excited states, or the degree of excitation of states is not clear.
\baselineskip=18pt

\centerline{ }
\centerline{ }
\centerline {{\bf 6. Perturbation Theory.}}
\centerline { }

\large {
\baselineskip=18pt

 We show now that in the considered model Dynamic Quantization
 can construct a mathematically correct PT in the fermion part of
 the Hamiltonian, exactly in the parameter \  {\mbox {\h
$e_a\,G\,\lambda_F^{-1}$}} \ .

\baselineskip=18pt

  The thing is that the available dimensionless parameters - the Lagrange
fields
 \  {\mbox {\h $e_a(x)$ }} \  -  can be made small enough
 {\it {indeed}} due to conditions (3.19). From these conditions and also
  formulae (2.14c), (3.10), (3.12), (4.6) and (5.7) we have the following
  estimation for the fermion part of the operator \  {\mbox {\h $H_{\phi}$}} \
 {\h {
 $$
 \Vert R_a^{\phi}\,\Vert \sim \Vert e_a\, \Vert \,N_0 \, \lambda_F^{-1}
  \eqno(6.1)
 $$
 }}
 Here \  {\mbox {\h $N_0$}} \ is the number of the fermion degrees of
 freedom in
the
  volume \  {\mbox {\h $V_e$}} \ , in which the field \  {\mbox {\h $e_a(x)$}}
\
 differs from zero:
 {\h {
 $$  N_0 \sim V_e \, \lambda_F^{-2}
 $$
 }}
 It is seen from estimation (6.1), that under sufficient numerical
 smallness of the
 fields \  {\mbox {\h $e_a(x)$}} \ and compactness of their support the
 expansion in the quantity
 \  {\mbox {\h $R^{\phi}_a$}} \ is mathematically correct.
  We state that under our assumptions one can establish the mathematic
correctness
  of Heisenberg equations (5.10) in a general form and solve these equations
 using the Euler method.
\footnote{ \large {
\baselineskip=14pt
 The problem of the calculation of the translation amplitudes
 with the change of the topology of the space is not considered here.}}.

\baselineskip=18pt
    Note that in the Feynman theory the reduction of the parameter
    \  {\mbox {\h $e_a$}} \ is useless in this sense since the order of
    the minimum wave length of the fermion fields \  {\mbox {\h $\lambda_F$}} \
  is equal to zero.

  In the process of the perturbation theory development the Heisenberg fields
can
  be expanded in a series in the constant \ {\mbox {\h $G$}}
{\h {
$$ A=A^{(0)}+A^{(1)}+\ldots ,
$$
\vspace{-3mm}\\
}}
where \ {\mbox {\h $A^{(0)}$}} \ is the zero approximation,
\ {\mbox {\h $A^{(1)}$}} \ is the first approximation, etc. It is evident that
the Dirac CR
 \ {\mbox {\h $[A^{(i)},\,B^{(j)}]^*$}} \ should be calculated in the
 \ {\mbox {\h $i+j$}}th approximation.

 Let \ {\mbox {\h ${\cal P}^i_a{ }^{(0)}(x)$}} \ and
 \ {\mbox  {\h  $\omega^a_i{  }^{(0)}(x)$}}  \  be  some  operator
fields, having
 in the zeroth approximation in the gravitational constant the following
 (nonzero) commutation relations:
{\h {
$$
 [\omega^{a(0)}_i(x),\,{\cal P}^{j(0)}_b(y)]^*=
 i\,\delta^j_i\,\delta^a_b\,\delta^{(2)}(x-y)   \eqno(6.2)
$$
\vspace{-3mm}\\
}}
  Take
\vspace{-15mm}\\
{\h {
$$   \psi_N(t_0,x)=\left(\,\gamma^0\,\Gamma^0(t_0)\right)^{-\frac{1}{2}}\,
  \kappa_N(t_0,x),     \eqno(6.3)
  $$
}}
\vspace{-3mm}\\
 where \ {\mbox {\h $\{\kappa_N(t_0,x)\}$}} \ is any complete set of numeral
 spinor fields satifying the conditions \ {\mbox {\h $\int
 d^2x \kappa^+_M\,\kappa_N=\delta_{MN}$}}.
 It is obvious that
 fields (6.3) satisfy conditions (3.12) at \ {\mbox {\h $t=t_0$}} \ .

The regularized Dirac field at the time \ {\mbox {\h $t_0$}} \
is expressed in the form
{\h {
$$   \psi^{(0)}(x)=\sum_{|N|<N_0}\,a_N\,\psi_N(t_0,x)          \eqno(6.4)
$$
\vspace{-1mm}\\
}}
The constant operators \ {\mbox {\h $\{a_N,a^+_N\}$}} \ satisfy the CR (3.6).

Extract the zeroth approximation from Eq.(5.10):
{\h {
$$  \dot{{\cal P}}^{i(0)}_a-\varepsilon_{ab}{ }^c\omega^b{\cal P}^{i(0)}_c=0
 $$
\vspace{-1mm}
$$
 \dot \omega^{a(0)}_i=
 \partial_i\,\omega^a-\varepsilon^a{ }_{bc}\,\omega^{b(0)}_i\,\omega^c ,
 $$
\vspace{-1mm}
 $$ \nabla_0\,\psi^{(0)}=0, \qquad \nabla_0\,\bar{\psi}^{(0)}=0   \eqno(6.5)
 $$
\vspace{-3mm}\\
 }}
Eqs.(6.5) are easily solved. Introduce the notations
{\h {
$$
\omega=\frac{1}{2}\,\omega_a\gamma^a, \ \qquad \ e=\frac{1}{2}\,e_a\gamma^a,
$$
\vspace{7mm}
$$ \omega_i=\frac{1}{2}\,\omega_{ai}\gamma^a, \ \qquad \ {\cal P}^i=
\frac{1}{2}\,{\cal P}^i_a\gamma^a                            \eqno(6.6)
$$
\vspace{-3mm}\\
}}
Let the c-number matrix field satisfy the equation
{\h {
$$
{\partial U}/{\partial t}=i\omega U, \ \ \ U(t_0,x)=1          \eqno(6.7)
$$
\vspace{-3mm}\\
}}
{}From group-theoretical arguments it is obvious that the operators
 \ {\mbox {\h $\chi_a$}} \ generate gauge transformations, which are easily
 eliminated. For this we introduce the fields with the tilde symbol:
{\h {
$$
 \omega_i=U\tilde{\omega}_i\bar{U}+i\,U\,\partial_i\bar{U},
 $$
\vspace{-3mm}
 $$ {\cal P}^i=U\tilde{{\cal P}}^i\bar{U}, \ \ \ \psi=U\tilde{\psi} ,
 $$
\vspace{-1mm}
 $$ \tilde{e}(t,x)=\bar{U}(t,x)\,e(t,x)\,U(t,x),   \eqno(6.8)
 $$
\vspace{-3mm}\\
 }}
\baselineskip=18pt
where \ {\mbox {\h $\bar{U}=\gamma^0U^+\gamma^0$}} .
Note that  \ {\mbox {\h $\bar{U}\,U=1$}} \ .
In the adopted notations, the solution of Eq.(6.5) is of the form
{\h {
$$ \tilde{\omega}^{(0)}_i(t)=\omega^{(0)}_i(t_0),
$$
\vspace{-3mm}
$$ \tilde{{\cal P}}^{i(0)}(t)={\cal P}^{i(0)}(t_0) ,
$$
\vspace{-1mm}
$$   \tilde{\psi}^{(0)}(t)=\psi^{(0)}(t_0)           \eqno(6.9)
$$
\vspace{-3mm}\\
}}
All the fields in (6.9) are taken at the point \ {\mbox {\h $x$}} .
The fields in (6.9) satisfy in the zeroth approximation not only equations
of motion but also Dirac CR (see (2.19) and Section 4).

\baselineskip=18pt

Let us write out the equations for the first-order corrections to the
Heisenberg fields (all the fields carry the tilde symbol, although,
to simplify the writing, it is absent in the formulae).
{\h {
$$ i\,\dot{\omega}^{a(1)}=\frac{1}{2}\,(8\pi G)\,\varepsilon_b{ }^{ca}e_c\cdot
$$
\vspace{-3mm}
$$ (\bar{\psi}\gamma^b\nabla_i\psi-\nabla_i\bar{\psi}\gamma^b\psi)+
$$
\vspace{-3mm}
$$ +\frac{1}{2}(8\pi G)^2\,\varepsilon_b{ }^{ca}\varepsilon_{ij}
\cdot(\bar{\psi}\,{\cal P}^j_c\,\gamma^b(\Gamma^0)^{-1}\Gamma^i\nabla_i\psi-
$$
\vspace{-3mm}
$$ -\nabla_i\bar{\psi}\,\Gamma^i\,(\Gamma^0)^{-1}
\gamma^b{\cal P}^j_c\psi),                     \eqno(6.10a)
$$
\vspace{-3mm}
$$   \dot{{\cal P}}^{i(1)}_a= -\frac{1}{8\pi
G}\,\varepsilon_{ij}(\partial_j\,e_a-
 \varepsilon_{ab}{ }^c\omega_j^be^c)-
 $$
\vspace{-1mm}
$$  -\frac{1}{2}\,(8\pi G)\,\varepsilon_a{ }^{bc}e_c
\bar{\psi}{\cal P}^i_b\psi \ ,                \eqno(6.10b)
$$
\vspace{-3mm}
$$  i\,\dot{\psi}^{(1)}=-i(\Gamma^0)^{-1}\Gamma^i\nabla_i\psi+
$$
\vspace{-3mm}
$$
  +\frac{i}{2}\,(8\pi G)\,\varepsilon_a{ }^{bc}e_c\,(\Gamma^0)^{-1}\gamma^a
   \psi\chi_b                                               \eqno(6.10c)
   $$
\vspace{-3mm}\\
   }}
In Eq.(6.10) all the fields in the right-hand side are taken in the zeroth
approximation according to (6.9).

\baselineskip=18pt
  One must also correct the Dirac CR (6.2) so that this CR will be valid
  to first-order inclusive in the gravitational constant.  This  problem
is solved
   in Section 4.

Thus, the chain of the regularized recursion equations does not differ formally
from
the analogous unregularized equations.
In essence however, in our theory a correctly defined
chain of recursion equations arises. This is  a  consequence  of
the regularization
of the Dirac field according to (6.4) and the replacement of the formal CR
(2.19) by Dirac CR.

 Finally, we consider one exactly solvable model. Let the system contain only
one fermion
 degree of freedom. This implies that from the set of modes
  \ {\mbox {\h $\{\psi_N\}$}} \ we choose mode
\ {\mbox {\h $\psi_0(x)$}} \ and impose the infinite series of constraints
(3.24) with \ {\mbox {\h $|N|>0$}} \ . Then, the fermi fields are of the form
{\h {
$$ \psi(x)=a_0\,\psi_0(x), \qquad \bar {\psi}(x)=\bar {\psi}_0(x)\,a^+_0,
$$
\vspace{-3mm}\\
}}
{}From this it follows that
\vspace{-10mm}\\
{\h {
$$ \psi(x)\,\psi(y)=0, \qquad \bar {\psi}(x)\,\bar {\psi}(y)=0   \eqno(6.11)
$$
\vspace{-3mm}\\
}}
  Relations (6.11) permit us to find the Dirac CR and solve the Heisenberg
  equations exactly. It is easy to see that for this purpose one more iterative
step
  is needed which will make Eq.(6.10b) more exact. It can be shown that
  as a result in the right-hand side
  of Eq.(6.10b) the symbol \ {\mbox {\h $\omega^b_j$}} \ denotes the sum
  \ {\mbox {\h $\omega^{b\,(0)}_j+\omega^{b\,(1)}_j$}} \ and everywhere
  in Eq.(6.10) symbols
  \ {\mbox {\h $\psi^{(0)} , \ \bar{\psi}^{(0)}, \
  \psi^{(1)} , \ \bar{\psi}^{(1)}$}} \  are replaced simply by symbols
  \ {\mbox {\h $\psi$}} \ and \ {\mbox {\h $\bar{\psi}$}} \ .
  In fact, subsequent iterations in gravitational constant lead to the
  corrections that contain the fields
  \ {\mbox {\h $\psi$}} \   or    \ {\mbox {\h $\bar{\psi}$}} \  to
  powers higher than the first. As a consequence of (6.11)
  all these corrections vanish. The terms containing the fields
  \ {\mbox {\h $\psi$}}
  (or  \ {\mbox {\h $\bar{\psi}$}} ) to the power two or higher
  vanish even in the case when these fields
   are separated by the operators
   \ {\mbox {\h $\omega^a_i$}} \  or \ {\mbox {\h ${\cal P}^i_a$}} .
   In this case one of the fields
  \ {\mbox {\h $\psi (\bar {\psi})$}} \
  must be moved so that it stands near the other field
  \ {\mbox {\h $\psi(\bar{\psi})$}} .  According to (6.11) such a term
  vanishes. In any case the commutators of the field
  \ {\mbox {\h $\psi (\bar {\psi})$}} \  with  the boson variables arising
  in the process of permutations
  increase the power of gravitational
  constant.
  It is clear therefore that by continuation of this process of
  permutations we shall obtain either zero or a term of a more higher degree
  in the gravitational constant.
  Thus, it can be seen that in the given case the expansion of the Heisenberg
  equations in the gravitational constant terminates on Eqs.(6.10).

  It is not difficult to see that in the case of one degree of
  freedom the expansions of the Dirac CR in the gravitational constant
  are also truncated.
\baselineskip=18pt

  The PT developed here justifies the Hypothesis A and B of Section 3.
\baselineskip=18pt
\centerline{ }
\centerline{ }
\centerline {{\bf 7. Conclusion. }}
\centerline { }

\baselineskip=18pt
    The results of [2] showed that the theory
    of the dynamic quantization developing here  is adequate
    for the construction
    of the quantum theory of gravity in a 2$+$1-dimensional space. The
constructed
    theory possesses the following necessary properties:
\vspace{3mm}\\
 a) The theory is unitary and causal; a set of evolution operators
 forms a group.\\
 b) The algebra of gauge transformations (5.12) is valid.\\
 c) A mathematically correct perturbation theory in the
 gravitational constant \ {\mbox {\h $G$}} exists.
 \vspace{-3mm}\\

  Here we suggest that a modification of dynamic quantization
  method makes the process of dynamic quantization
  in (3$+$1)-dimensional space-time easier. The main difference between that
theory
  and the present one is that it is necessary to identify not only fermion
  but also boson gauge invariant creation and annihilation operators.
  We think  that it makes sense to study the supersymmetric
variant of the theory.
  In supersymmetric theories the filling
of the vacuum can be realized in such a way
  that the contributions of the boson and fermion zeroth oscillations to the
  energy-momentum tensor of matter cancel out. Probably, in this case a PT
  in the gravitational constant analogous to the PT considered in Section 6
  is to be valid.

\baselineskip=18pt
\centerline{ }
\centerline{ }
\centerline {{\bf Appendix. }}
\centerline { }

\baselineskip=18pt
  Represent the operators \ {\mbox {\h $ \{a^+_N,a_N\}$}} \, ,
  vectors (3.3) and scalar product in a natural way. Let
   \ {\mbox {\h $ \alpha_N$}} \ and \ {\mbox {\h $ \bar {\alpha}_N$}} \ be
   the elements of a Grassmann algebra over complex numbers.
   Then
   {\h {
$$
\vert \, N,0 \rangle =1, \qquad \vert \, N,1 \rangle=\bar {\alpha}_N
\eqno(A1)
$$
}}
and any fermion state is represented by the function on the
variables
 \ {\mbox {\h $ \{\bar {\alpha}_N\}$}} \ .
 By this the action of creation and annihilation operators looks like
 {\h {
 $$
 a^+_N\,F\{\bar {\alpha}\}=\bar {\alpha}_N\,F\{\bar {\alpha}\}, \qquad
 a_N\, F\{\bar {\alpha}\}=
 {\partial}_{\bar {\alpha}_N}\,F\{\bar {\alpha}\}  \eqno(A2)
$$
}}
The function conjugated to
\vspace{-12mm}\\
{\h {
$$
  F_{\Lambda}(\bar {\alpha}) = \prod_N (\Lambda_{0\,N}+\Lambda_{1\,N}\,
  \bar {\alpha}_N)\cdot
  $$
  }}
  is of the form
\vspace{-13mm}\\
{\h {
$$
  \bigl( F_{\Lambda}(\bar {\alpha})\bigr)^+
  = \bar {\prod_N} (\bar {\Lambda}_{0\,N}+\bar {\Lambda}_{1\,N}
  \alpha_N)
  $$
  }}
Here symbol  \ {\mbox {\h $\bar {\prod}$}} \ denotes the product
in inverse order with comparison  the product in previous formula,
  \ {\mbox {\h $ \Lambda_N$}} \ and \ {\mbox {\h $ \bar {\Lambda}_N$}} \
  are mutually-conjugated complex numbers.
  The scalar product (3.5) is expressed by the formula
\vspace{-13mm}\\
  {\h {
  $$
  \langle \Lambda \vert \, \Sigma \rangle = \int \bigl( F_{\Lambda}(\bar
{\alpha})
  \bigr)^+\,F_{\Sigma}(\bar {\alpha})\,\prod_N e^{-\bar {\alpha}_N\alpha_N}
  d\bar {\alpha}_N\,d\alpha_N                 \eqno(A3)
  $$
  }}

 In a field variant the set of Grassmann elements \ {\mbox {\h $\{\alpha_N\}$}}
 \ and \ {\mbox {\h $\{{\bar {\alpha}}_N\}$}} \ correspond to Grassmann
 spinor fields \ {\mbox {\h $\lambda(x)$}} \ and
 \ {\mbox {\h $\bar {\lambda}(x)$}} \ . The fermion wave functions
   \ {\mbox {\h $F_{\Lambda}(\bar {\lambda})$}} \ are the functionals on
   \ {\mbox {\h $\bar {\lambda}(x)$}} \ ,  and their conjugated
   \ {\mbox {\h $F^+_{\Lambda}$}} \ are the functionals on
   \ {\mbox {\h $\lambda(x)$}} \ .
  The action of operators \ {\mbox {\h $\bar {\psi}$}} \ and
   \ {\mbox {\h $\psi$}} \ is defined as:
{\h {
$$
   \bar {\psi}(x)\,F(\bar {\lambda})=\bar {\lambda}(x)\,F(\bar {\lambda}),
$$
$$
  \psi(x)\,F(\bar {\lambda})=
  (\Gamma^0)^{-1}(x)\,\bigl(\frac{\delta\,F(\bar {\lambda})}
  {\delta\,\bar {\lambda}(x)}\bigr)
$$
}}
  Let the fields
   \ {\mbox {\h $\lambda(x)$}} \  and \ {\mbox {\h $\bar {\lambda}(x)$}}
   satisfy the same equations as the fields
   \ {\mbox {\h $\psi(x)$}} \ and  \ {\mbox {\h $\bar {\psi}(x)$}} \
   correspondingly.
 Then the expression
\ {\mbox {\h $ \int\,d^2x\,\bar{\lambda}\,\Gamma^0\lambda$}} \ is
gauge invariant and the
functional measure on the phase space of fermion fields is expressed
in the form:
\vspace{-12mm}\\
{\h {
$$
  exp(-\int\,d^2x \,\bar{\lambda}\,\Gamma^0\lambda)\,\prod_x\,
  \bigl( det\Gamma^0(x)
  \bigr)^{-1}\,d\bar {\lambda}(x)\,d\lambda(x)               \eqno(A4)
$$
}}
  In selected variables the measure (A4)
transforms to measure from integrals (A3). To see this let expand the fields
   \ {\mbox {\h $\lambda$}} \  and \ {\mbox {\h $\bar {\lambda}$}}
   analogously to (3.10) \\
(\ {\mbox {\h $\lambda(x)=\sum_N\,\alpha_N\,\psi_N(x)$}} \ ) and then use
the relations (3.12).

\baselineskip=18pt
The measure in (A3) is
preferable since it can be regularized by dynamically invariant way:
\vspace{-13mm}\\
{\h {
$$
  ({\cal D}\bar {\psi}\,{\cal D}\psi)_{Reg}=\prod_N{ }^{\prime}
  e^{-\bar {\alpha}_N\alpha_N}\,d\bar {\alpha}_N\,d\alpha_N   \eqno(A5)
$$
}}
Here the symbol \ {\mbox {\h $\prod_N^{\prime}$}} \ means the product
over some finite set of indices \ {\mbox {\h $N$}} \ .

\baselineskip=18pt

      Compare the measure (A5) with fermiom measure which is used for
calculations of axial anomaly at Euclidean formulation of the gauge theory.
Let us consider in D-dimensional Euclidean space the Dirac field in
external gauge field \ {\mbox {\h $A_i(x)$}} \, .
Let \ {\mbox {\h $\{\psi_N\}$}} \ be the complete orthogonal set
of eigenfunctions of Dirac operator:
     {\h {
     $$
      -i\gamma^i\,\nabla_i\psi_N=\varepsilon_n\,\psi_N
$$
}}
Here \ {\mbox {\h $\nabla_i=\partial_i-i\,A_i, \  \gamma^i$}} \ are Dirac
matrices in Euclidean space. Expand the Dirac fields in the set
 \ {\mbox {\h $\{\psi_N\}$}}:
 {\h {
 $$ \psi(x)=\sum_N \alpha_N\,\psi_N(x), \qquad
   \bar{\psi}(x)=\sum_N \bar {\alpha}_N\,\bar{\psi}_N(x)
 $$
 }}
 (Compare with (3.10)). Then the fermion measure
 \ {\mbox {\h ${\cal D}\bar {\psi}\,{\cal D}\psi$}} \
 can be formally expressed as
  \  {\mbox  {\h  $\prod_N  d\bar   {\alpha}_N\,d\alpha_N$}}   and
regularized one
  according to
\vspace{-14mm}\\
{\h {
$$
  ({\cal D}\bar {\psi}\,{\cal D}\psi)_{Reg}=
  \prod_{\{N :\,\vert \, \varepsilon_N \vert \, <
  \Lambda\}} d\bar {\alpha}_N\,d\alpha_N, \qquad
 \Lambda \longrightarrow \infty  \eqno(A6)
$$
}}
\baselineskip=18pt
 (Compare with (A5)). Since the variables
 \ {\mbox {\h $\{\bar {\alpha}_N,\alpha_N\}$}} \ are gauge invariant so the
 measure (A6) is also gauge invariant. However as it has been shown in
 Dissertation of the author [7] the measure
 (A6) contains the axial anomaly.
\footnote{ \large {
\baselineskip=14pt
 The same result has been obtained by Fujikawa [8]}}
 By our opinion the dynamic invariant measures of the kind (A5) differ
 principally from Euclidean measures of the kind (A6) by that the first
 one do not contain any gauge anomalies.

\centerline{ }
\centerline{ }

\centerline{ {\bf References}}
\vspace{3mm}

\baselineskip=18pt

1. S.N. Vergeles, {\underbar {JETP Lett.}} {\bf 49} (1989) 375;
 {\underbar {JETP}}. {\bf 96} (1989) 445;
 {\underbar {Yad.Phys.}} {\bf 50} (1989) 1476;
 {\underbar {Int.J.Mod.Phys.A.}} {\bf 5} (1990) 2117.}}\\
2. S.N.Vergeles, {\underbar {JETP.}} {\bf 102} (1992) 1739.\\
3. V.N. Gribov,  Preprint KFKI-66. Budapest. 1981.\\
4. S.N.Vergeles, {\underbar {JETP.}} {\bf 95} (1989) 397.\\
5. E. Witten, {\underbar {Nucl.Phys.}} {\bf B311} (1988/89) 46;
 {\bf B323} (1989) 113.}\\
6. P.A.N.Dirac, {\underbar {Lectures on Quantum Mechanics }}
(Yeshiva University, New York, 1964) \\
7. S.N.Vergeles, {\underbar {Some problems of gauge fields theory}}
(Dissertation, Chernogolovka, Moskow District 1979);
see in A.A. Migdal, {\underbar {Phys.Lett.} {\bf 81B} (1979) 37. \\
8. K. Fujikawa, {\underbar {Phys.Rev.Lett.}} {\bf 42} (1979) 1195
}

\baselineskip=18pt

\end{document}